\documentclass[conference]{IEEEtran}
\IEEEoverridecommandlockouts
\usepackage{cite}
\usepackage{amsmath,amssymb,amsfonts}
\usepackage{graphicx}
\usepackage{textcomp}
\usepackage{xcolor}

\usepackage{amsmath}
\usepackage{amsthm}
\usepackage{listings}
\usepackage{multirow}
\usepackage{epsfig}
\usepackage{endnotes}
\usepackage{xspace}
\usepackage{algorithm}
\usepackage[noend]{algpseudocode}
\usepackage{soul}
\usepackage{balance}
\usepackage{color}
\usepackage{enumitem}
\usepackage{url}

\setlist[itemize]{leftmargin=15pt}
\definecolor{pqRed}{RGB}{255,0,102}

\newcommand{\scaleGANN}{ScaleGANN\xspace}

\usepackage{etoolbox}
\newbool{inline_remark}
\setbool{inline_remark}{false} 
\newcommand{\tl}[1]{%
  \ifbool{inline_remark}
  {%
    {\color{blue}[tao: #1]}%
  }
  {}%
}

\newcommand{\stitle}[1]{\vspace*{0.3em}\noindent{\bf #1.\/}}

\def\BibTeX{{\rm B\kern-.05em{\sc i\kern-.025em b}\kern-.08em
    T\kern-.1667em\lower.7ex\hbox{E}\kern-.125emX}}

\begin{document}
\title{ScaleGANN: Accelerate Large-Scale ANN Indexing by Cost-effective Cloud GPUs}

\author{
\IEEEauthorblockN{Lan Lu}
\IEEEauthorblockA{
University of Pennsylvania \\
lanlu@seas.upenn.edu
}
\and
\IEEEauthorblockN{Peiqi Yin}
\IEEEauthorblockA{
CUHK \\
pqyin22@cse.cuhk.edu.hk
}
\and
\IEEEauthorblockN{Isaac Yang}
\IEEEauthorblockA{
Duke Kunshan University \\
isaac.y@dukekunshan.edu.cn
}
\and
\IEEEauthorblockN{Tao Luo}
\IEEEauthorblockA{
University of Pennsylvania \\
taoluo71@seas.upenn.edu
}
\and
\IEEEauthorblockN{Hua Fan}
\IEEEauthorblockA{
Alibaba Cloud \\
guanming.fh@alibaba-inc.com
}
\and
\IEEEauthorblockN{Wenchao Zhou}
\IEEEauthorblockA{
Alibaba Cloud \\
zwc231487@alibaba-inc.com
}
\and
\IEEEauthorblockN{Feifei Li}
\IEEEauthorblockA{
Alibaba Cloud \\
lifeifei@alibaba-inc.com
}
\and
\IEEEauthorblockN{Boon Thau Loo}
\IEEEauthorblockA{
University of Pennsylvania \\
boonloo@seas.upenn.edu
}
}

\maketitle

\begin{abstract}
    Graph-based ANNS algorithms have gained increasing research interest and market adoption due to their efficiency and accuracy in retrieval. Existing approaches primarily rely on CPUs for graph index construction and retrieval, but this often requires significant time, especially for large-scale and high-dimensional datasets. Some studies have explored GPU-based solutions. However, GPUs are costly and their limited memory makes handling large datasets challenging. 
    In this paper, we propose a novel end-to-end system \scaleGANN that enables users to efficiently construct graph indexes for large-scale, high-dimensional datasets by leveraging low-cost spot GPU resources in a distributed cloud system. \scaleGANN utilized the idea of divide-and-merge, with an optimized vector partitioning algorithm to further improve the indexing time and space efficiency while guaranteeing good index quality. Its novel resource allocation strategy realized multi-GPU indexing parallelism and overall cost-effectiveness for both build and query. Besides, we designed a task scheduler and cost model for better spot instance management and evaluation.
    We tested our system on large real-world datasets. Experiment results show that our approach can significantly accelerate the index build time to up to 9x times at even 6x lower price compared with the state-of-the-art extendable ANNS benchmark DiskANN, while we preserve the same scalability and similar search quality as DiskANN.
\end{abstract}


\section{Introduction}
\label{sec:introduction}

Approximate Nearest Neighbor Search (ANNS) is a fundamental technique in vector databases. 
By building an index over the dataset and limiting the search to vectors similar to a given query, ANNS significantly enhances retrieval speed, with a minor sacrifice in accuracy. It has been widely adopted across multiple domains including information retrieval~\cite{flickner1995query, wang2021comprehensive}, pattern recognition~\cite{cover1967nearest, li2019approximate}, recommendation systems~\cite{covington2016deep, chen2022approximate}, and machine learning~\cite{liu2024retrievalattention, li2023skillgpt}. 
Among various types of index, graph-based indexes~\cite{wang2021comprehensive, fu2017fast, malkov2018efficient, ootomo2024cagra, malkov2014approximate} generally achieve better search quality~\cite{fu2017fast, li2019approximate}
due to fine granularity graph traversal. Therefore, they are widely adopted in practice.

\stitle{Scaling graph indexing to large datasets}
An increasing number of applications operate over large-scale datasets, posing new challenges for ANNS. For example, e-commerce platforms and social medias routinely store data at the billion-scale~\cite{arthurs2018researching, schuhmann2022laion}, often with thousands of dimensions or even higher. Many of these services demand frequent index updates and reconstructions to maintain accurate query results, such as web search engines and recommendation systems~\cite{Shi2025ScalableOG}.
Yet, efficiently constructing graph indexes for large datasets remains challenging due to limited storage, and the time-consuming extensive distance computations required for edge discovery. For billion-scale, high-dimensional datasets, index construction can take multiple terabytes of memory, and last for several days to weeks to complete~\cite{jayaram2019diskann, wang2024starling}. 

\stitle{Partition-based scalability and problems}
The typical solution to graph indexing with high storage demands is data sharding, while trading off build and query performance. Incorporated with disk storage, DiskANN~\cite{jayaram2019diskann} replicates each vector across shards to maintain global connectivity and facilitates shard index merging after the subgraphs are built. 
However, replication increases I/O and lengthens build-time roughly in proportion to the number of vectors replicated~\cite{CVPR_tutorial, ootomo2024cagra}.
In contrast, other approaches split the dataset into independent shards without replication or merging, such as GGNN~\cite{groh2022ggnn}.
Yet, during querying, each shard must be searched independently, increasing latency and compute.

\stitle{GPU-based accelerations and challenges}
To accelerate index building, existing works proposed GPU-based implementations~\cite{zhao2020song, ootomo2024cagra, yu2022gpu, groh2022ggnn} in construction and querying. 
However, limited GPU VRAM, frequent GPU-CPU transfers, and higher on-demand pricing make them hard to use cost-effectively at scale.
According to AWS~\cite{awsspot} and Alibaba cloud~\cite{alispot} pricing, adding four 16GB V100 GPUs to a machine with 32 vCPUs and 256 GiB RAM can increase the cost by 6–7×. The costs can escalate with higher memory capacity, greater computational performance, or additional GPUs.

\stitle{Balancing index build and search performance}
Besides, existing ANNS systems run entirely on CPUs or GPUs for both index construction and query. The former slows index construction, while the latter increases serving latency relative to CPU-based ones~\cite{liu2024retrievalattention, chen2024memanns}.
Especially, GPU-based shard querying for large datasets incurs additional distance computations and frequent GPU-CPU data transfers. Furthermore, the unpredictable distribution and frequency of queries in real-world applications often limit batching and parallelism opportunities, leading to expensive but suboptimal GPU utilization during search.
Most notably, CPU-based querying throughput meets the performance requirements of most applications~\cite{liu2024retrievalattention, chen2024memanns}, making GPU-based querying unnecessary in many cases.

\stitle{Our solution: \scaleGANN}
Therefore, we present \textbf{\scaleGANN}, a novel GPU-CPU hybrid and cloud-native framework that enables efficient and disk-resident scalable graph ANNS indexing for large-scale, high-dimensional datasets, which also balances cost and search performance.

\scaleGANN propose use of cloud GPUs (especially cost-effective spot instances up to 10x cheaper~\cite{awsspot, alispot, miao2024spotserve}) exclusively for accelerating the one-time (or periodic) index construction, while delegating long-running, latency-sensitive query serving~\cite{Shi2025ScalableOG} to CPUs. 
This design choice reflects the nature of the workload: index construction is typically a one-time, compute-intensive process, while query serving is long-running and highly cost-sensitive. This also enables resource isolation between build and query, ensuring stable query performance, while allowing the built index to be replicated for scalable serving across multiple machines.

At its core, \scaleGANN extends a disk-incorporated partition-and-merge design with enhanced selective replication to reduce disk I/Os, memory usage and build time, while maintaining global graph connectivity and quality. Unlike DiskANN which duplicates each point at least once with a uniform count across the dataset, \scaleGANN effectively replicates only the most essential vectors based on our proposed pruning techniques. 
Once obtaining the partitioned vector shards,
\scaleGANN's decoupled indexing tasks are further accelerated by multi-GPU parallelism.
We also parallel the vector assignment to shards, and design a disk buffer state check during index merge to handle the non-deterministic vector order within each shard caused by thread parallelism.

To address the inherent instability and evaluate the cost-effectiveness of preemptive spot instances, we introduce a spot instance task scheduler and a intuitive cost model. 
We specify spot GPU instance use only for shard indexing tasks which are parallel and small in size limited by GPU memory limit, and leave the tightly-coupled partition and merge on non-spot machines. 
Then we deploy a spot instance task scheduler on CPUs to manage the assignment of each shard indexing task to an available GPU spot instance. By estimating task execution time which is proportional to the shard size, the scheduler tries to allow task assignment only to spot instances with sufficient remaining availability.
Lastly, we design a intuitive cost model for evaluating the expense of different resources, and illustrate the cost-effectiveness of spot instance with an example in experiments.

While \scaleGANN's general framework allows the integration with diverse indexing algorithms, in this work, we integrate it with the state-of-the-art GPU-based CAGRA~\cite{ootomo2024cagra} algorithm, and demonstrate its scalability, performance, retrieval quality, and cost-effectiveness on real-world datasets ranging from millions to billions of vectors.

Notably, \scaleGANN significantly reduces index construction time relative to disk-incorporated CPU‑based baselines, while maintaining similar search quality at matched recall.
On high‑dimensional Laion100M dataset and index build degree $R$=128, we observe up to 9× overall acceleration over DiskANN.
Besides, compared to GPU-based approach, \scaleGANN realizes affordable indexing time for large datasets, while the partition-merge-based \scaleGANN achieves $\sim$3× acceleration in search latency at same recall compared to the partition-based Extended CAGRA (we extend CAGRA for large datasets) and GGNN.
While \scaleGANN achieves even $\sim$3× overall build acceleration over GGNN on Laion100M ($R$=128), its GPU build‑only time (without partition and merge time) on the replicated dataset always remains $<$2× that of Extended CAGRA, benefiting from our enhanced selective vector replication for both the build efficiency and search quality.
Importantly, these improvements can be achieved at much lower dollar cost (up to 6x cheaper than DiskANN) if using GPU spot instances.

In summary, our main contributions are as follows:
\begin{itemize}
    \item We propose the first \textbf{end-to-end} spot-instance-aware GPU system for graph ANNS indexing, which is highly efficient, scalable, and extendable to cost-effective cloud resources. 
    \item We design a novel resource allocation strategy which supports multi-GPU parallelism for decoupled indexing tasks and CPU-based querying, to balance build efficiency, search latency and cost-effectiveness.
    \item We improve the partition-merge-based construction approach especially with selective vector replication and parallelism to save indexing time and memory usage while ensuring global connectivity. Search performance remains unaffected or even improved after replication pruning. 
    \item We design a spot instance scheduler and cost model to manage task assignment and analyze the cost-effectiveness.
   \item We integrate \scaleGANN with CAGRA, and validate it on real-world, large-scale high-dimensional datasets.
    Compared to DiskANN, \scaleGANN brings a significant acceleration in index build at a considerable low expense, while maintaining similar high search performance.
\end{itemize}

\section{Background}

\subsection{Graph-based ANNS Index}
Among various ANNS index structures~\cite{douze2024faiss,abbasifard2014survey,li2019approximate, liu2004investigation}, the graph-based index~\cite{fu2017fast,malkov2014approximate, malkov2018efficient,wang2021comprehensive} is considered the most effective due to its high accuracy and low computational cost. It builds a proximity graph~\cite{wang2021comprehensive, fu2017fast, malkov2018efficient} over the vector dataset, typically using k-nearest-neighbor~\cite{fu2016efanna, hajebi2011fast, ootomo2024cagra} or small world graphs~\cite{malkov2014approximate, malkov2018efficient}
, where each node represents a vector and is connected to its similar vectors. 
Existing graph-based indexes use different construction techniques, such as NSG~\cite{fu2017fast} and HNSW~\cite{malkov2018efficient}, to balance the trade-off between index quality and construction overhead. 

Additionally, SONG ~\cite{zhao2020song}, CAGRA~\cite{ootomo2024cagra}, and GANNS~\cite{yu2022gpu} leverage GPUs to speed up the graph construction. The extensive distance calculations between nodes and their neighbors can be efficiently parallelized by GPU using matrix multiplication (\texttt{matmul}). While CAGRA~\cite{ootomo2024cagra} achieves 2.2–27× speedups compared to leading CPU-based methods such as HNSW~\cite{malkov2018efficient}, all these works currently support only datasets that fit entirely within a single GPU's memory.

Vamana, introduced by DiskANN\cite{jayaram2019diskann}, enables graph construction at billion-scale through a partition-and-merge approach. It begins by dividing the database into multiple shards using K-means clustering. To maintain overall graph connectivity, each node is assigned to multiple shards (typically two), serving as a link between partitions. 
Each subgraph construction is then sequentially executed and parallelized by multiple CPU threads, enabling independent and efficient subgraph construction. 
Once the individual subgraphs are built, they are merged into a single, connected graph over the complete dataset via edge union, preserving the local structure within shards while ensuring global connectivity. 
By leveraging disk storage to overcome memory constraints during partition and merge, DiskANN enables support for billion-scale datasets.

GGNN~\cite{groh2022ggnn} partitions the graph into smaller shards, assigning each GPU to construct the graph for a single shard. Unlike Vamana, which partitions before construction and merges after, GGNN performs partitioning during graph construction and merging during the search phase. Its partitioning strategy avoids node duplication, since each node appears in only one subgraph and each subgraph resides independently in GPU memory. During similarity search, GGNN exploits GPU parallelism by having each GPU search for the local top-$k$ results within its subgraph. These local results are then aggregated across GPUs to compute the final top-$k$ results.


\subsection{Spot Instance}
Modern cloud providers offer spot GPU instance products (e.g., Alibaba Cloud ECS~\cite{alispot}, Azure spot VMs~\cite{azurevm} and AWS spot instances~\cite{awsspot}), with a much cheaper price compared to on-demand instances (up to 90\% price reduction). 
However, these spot instances are preemptible, and cloud providers may terminate the service at any time. When the spot service is terminated or preempted by other on-demand instances, the user will generally receive a notification ahead (i.e., 5 minutes by Alibaba ECS~\cite{alispot}). Users can migrate their working jobs during that period. Besides, most spot instances have a safe duration time at initialization (i.e., 1 hour~\cite {alispot}), while spot instances without such a protection period are even cheaper.

\section{Motivation}
\label{sec:motivation}

As large-scale ANNS applications become prevalent, efficient graph index construction is ever more critical. In this section, we present several case studies to analyze the bottlenecks in graph indexing for large datasets and explore potential optimization directions, thereby motivating the design of our framework. We conduct a simple sampling analysis on the Sift1B and Laion1B datasets. The dimensionality and data types of these datasets are summarized in Table~\ref{tab:index-build-time-comparison}. The unit of time is second (s) in this section.

\begin{table}[t]
    \caption{Time Breakdown (s) of DiskANN Index Construction.}
    \label{tab:sift100M-time-breakdown(DiskANN)}
    \vspace{-0.5em}
    \centering
    \footnotesize
    \resizebox{1\linewidth}{!}{
    \begin{tabular}{l|l|l|l}
    \hline
      & \textbf{Disk Partition} & \textbf{Index Build on shards} & \textbf{Disk Merge} \\
     \hline
     \textbf{Sift100M} & 129.1 & 3021.8 & 978.5 \\
     (R=32, L=64) & & & \\
     \hline     
     \textbf{Sift100M} & 157.5 & 7847.9 & 1591.8 \\
     (R=64, L=128) & & & \\
     \hline
    \end{tabular}}
    \vspace{-0.5em}
\end{table}

\noindent \textbf{Slow Graph Index Construction on Large Datasets, and Shard Index Build Dominates the Time.}
Table ~\ref{tab:sift100M-time-breakdown(DiskANN)} presents the time breakdown of graph index construction on the Sift100M dataset using DiskANN. Given a memory cap of 16 GB, DiskANN performs partitioned index construction and merging. We conduct experiments using 80 parallel CPU threads, with two different configurations of graph construction parameters where final index degree $R$ and intermediate graph degree $L$ are set to (32, 64) and (64, 128), respectively.
We observe that DiskANN takes a considerable amount of time to complete index construction on Sift100M. As the parameters $L$ and 
$T$ increase, the construction time becomes even longer. Furthermore, as shown in Table ~\ref{tab:index-build-time-comparison}, the Sift dataset is relatively simple compared to others such as Laion. When the same number of vectors are sampled from different datasets, we find that higher data dimensionality and a shift from integer to floating-point data types significantly increase DiskANN’s index construction time. This is primarily due to the larger number and higher cost of distance computations in high-dimensional spaces with floating-point operations.
In addition, we observe that index building on data shards is the dominant time contributor, compared to data partitioning and disk merging. As the parameters $L$ and 
$R$ increase, the proportion of time spent on index building becomes even larger—again, due to the increasing cost of distance computations.
These findings clearly demonstrate the necessity of accelerating the graph index build process, especially for large-scale datasets.

\begin{table}[t]
    \caption{Index Build Time (s) of CAGRA and DiskANN.}
    \label{tab:index-build-time-comparison}
    \vspace{-0.5em}
    \centering
    \footnotesize
    \resizebox{1\linewidth}{!}{
    \begin{tabular}{l|l|l|l|l}
    \hline
     \textbf{Dataset} & \textbf{Dimension} & \textbf{Data Type} & \textbf{CAGRA} & \textbf{DiskANN} \\
      & & & (R=32, L=64) & (R=32, L=64) \\
     \hline
     \textbf{Sift1M} & 128 & uint8 & 12.7 & 13.2 \\
     \hline
     \textbf{Laion1M} & 768 & float & 25.8 & 112.4 \\
     \hline
\end{tabular}}
\vspace{-0.5em}
\end{table}

\noindent \textbf{GPU Accelerates the Index Build.}
Table ~\ref{tab:index-build-time-comparison} compares the graph index construction time between the GPU-based CAGRA algorithm and the CPU-based DiskANN algorithm on the Sift1M and Laion1M datasets. Due to GPU memory limitations (typically 16–80 GB), CAGRA can only support small-scale datasets for index construction. Therefore, we restrict the comparison to 1M-scale datasets.
In this setting, DiskANN is able to directly construct the graph index in memory without partitioning and merging, and so can CAGRA. Thus, both methods operate under the same assumption that there is no need for partitioned indexing or post-processing merge steps.
We conduct experiments using an NVIDIA V100 GPU with 16 GB memory and 5120 threads. During index construction, we fix the final index degree $R$ and the intermediate graph degree $L$ to 32 and 64, respectively.
Our results show that, compared to CPU-based DiskANN, GPU-based CAGRA significantly accelerates the index build process for datasets with higher dimensionality and floating-point data types, where distance computations are denser and more computationally expensive.
These observations confirm the feasibility and effectiveness of using GPUs to accelerate graph index construction for large datasets with high dimensionality.

\noindent \textbf{Scalable Datasets Brings Storage Limitation.}
Large-scale and high-dimensional datasets can span several or more terabytes~\cite{laion5B}, while graph index itself introduces additional storage overhead. In contrast, CPU and GPU memory are limited: modern GPUs often offer only 16–80GB, far below the requirements for index construction. Besides, the partition-and-merge based approaches like DiskANN further tighten this constraint by uniformly duplicating every vector.

\noindent \textbf{GPU Querying Leads to High Latency and Expense.\tl{this subtitle seems not accurate, should be "Query using GPU is inefficient both in latency and Cost/expense"}}
Limited accelerator memory forces GPU-based querying to operate over partitioned index subgraphs. 
In GGNN, shards are queried independently with subsequent sorting and merging, whereas swap-based approaches induce frequent host–device transfers. These introduce high search latency. 
Moreover, unlike the one-time cost of index construction, production-level querying necessitates maintaining GPU availability for stochastic arrivals, which is typically cost-inefficient and yields suboptimal resource utilization relative to CPU-based deployments.

Based on the above findings, we motivate both the necessity and feasibility of using GPUs to accelerate graph index construction for large-scale datasets, while also highlighting why GPUs are not suitable for index querying.
In response, we propose a novel framework that accelerates graph index construction using GPU resources under a partition-and-merge paradigm, while relying on CPUs to perform index querying. This design strikes a balance between efficiency and cost-effectiveness, leveraging the parallelism of GPUs for intensive index building tasks, and the flexibility and scalability of CPUs for handling dynamic, on-demand query workloads.
\section{Framework of \scaleGANN}
\label{sec:framework}
\begin{figure}[t]
    \centering
    \includegraphics[width=0.48\textwidth]{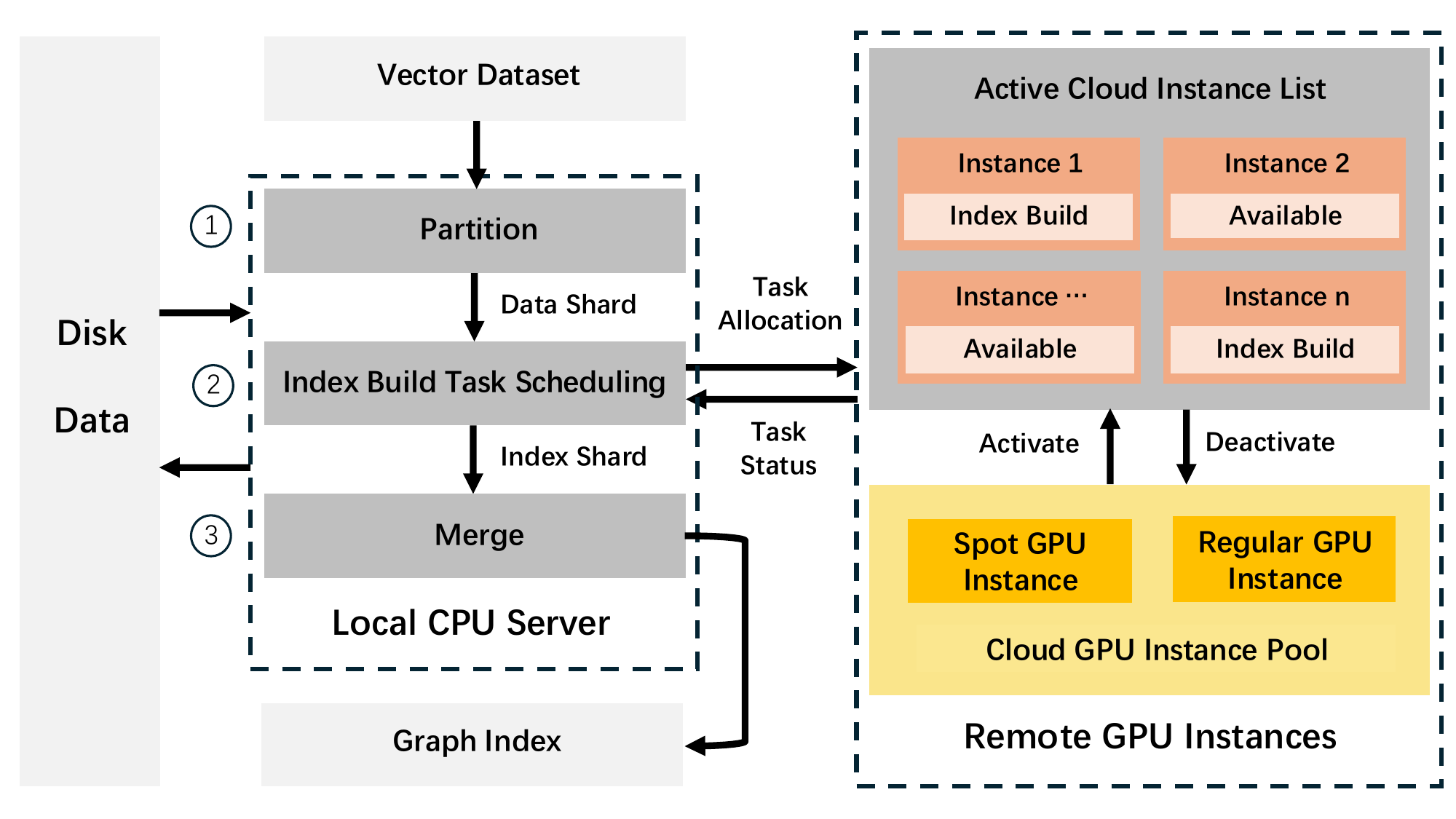}
    \caption{Illustration of \scaleGANN's build Framework.}
    \label{fig:scaleGANN}
\end{figure}

We illustrate the overview of \scaleGANN's build framework in Figure~\ref{fig:scaleGANN}. 
Given an input vector dataset, \scaleGANN uses two types of computational resources, local CPU resources and remote cloud GPU instances, to construct the output merged index. 
In particular, \scaleGANN prefers GPU spot instances when available rather than regular GPU services for better cost-effectiveness.
It also implements a cloud instance task scheduler to manage task-instance assignment, and disk-supported data handling for memory efficiency.

\noindent \textbf{Workflow:} The end-to-end index construction workflow consists of three key stages: data partitioning, shard-wise index construction using GPU instances, and index merging. 
(1) Data Partitioning: Given a large input dataset necessary for partitioning, we apply k-means clustering and assign vectors to data shards. Vectors are selectively replicated and wisely assigned across shards to preserve global connectivity and shard locality, while reducing memory usage and processing overhead. We also realize parallel implementation for partition efficiency. The detailed partitioning mechanism is presented in Sec~\ref{sec:partition}.
(2) Index Build: We use remote GPU instances to construct indexes for data shards, leveraging parallel execution to speed up the distance computations. Each available GPU instance is assigned an independent shard-level indexing task, and invokes a GPU-based indexing algorithm such as CAGRA. 
Since each per-shard subgraph is built independently, no back-forth CPU-CPU communication or inter-GPU communication is necessary during index building.
This design not only eliminates redundant CPU-GPU data transfers, but also enables efficient parallelism of multiple idle GPUs for accelerating the otherwise time-consuming index construction process.
We also implement a task scheduler for task assignments to appropriate cloud GPU instances, which will be described later.
(3) Index Merging: Finally, constructed shard-level indexes are merged into a unified global index, using replicated vectors across shards as connections, as discussed in prior work~\cite{jayaram2019diskann}.

\noindent \textbf{Resource Allocation:} Our framework utilizes two types of computational resources: local CPUs and remote GPUs. We employ local CPU resources not only for querying, but also for the tightly-coupled data partitioning and index merging. 
Meanwhile, the remote cloud GPU resources, in particular idle spot instances, are used to accelerate the decoupled shard-level indexing with intensive distance computation.

Although certain steps in partitioning and merging could theoretically benefit from GPU acceleration, we deliberately choose to execute them on CPUs for two main reasons:
1. Non-bottleneck nature: Compared to index construction, the time spent on partitioning and merging does not constitute the system’s performance bottleneck. Additionally, transferring data between CPU and GPU memory introduces extra overhead that may offset any potential acceleration benefits.
2. Disk-intensive and highly-coupled logic: Both partitioning and merging involve frequent interactions with disk storage and contain substantial non-parallel, tightly-coupled logic beyond distance computations. These characteristics make them less suitable for GPU acceleration, which is optimized specifically for large-scale parallel numeric operations.
By reserving GPU resources for the most computation-intensive index construction, we achieve an effective balance between performance and resource utilization.

\noindent \textbf{Cloud Instance Task Scheduler:}
When constructing shard-wise indexes using cloud GPU instances, we require a task scheduler to assign appropriate index construction tasks to available GPU instances. 
Specifically, the scheduler needs to maintain two key components: A task list that tracks all pending index construction tasks for data shards, and a cloud instance list that manages active remote GPU instances especially the spot GPU instances and records their status. The cloud instance list records the following statuses for each GPU instance: (1) Active: A GPU instance that has been successfully rented from cloud service provider's instance pool is marked as active and added to the instance list. Conversely, if the instance service is terminated, it is marked as inactive and removed from the list. (2) Available: Within the cloud instance list, if a GPU is currently executing a task, it is marked as unavailable; otherwise, it is available. (3) Time remaining: For each GPU instance especially the spot instances, if we have accurate information about its remaining active lifetime, we record it accordingly.
Note that \scaleGANN always prefers activating the spot GPU instances at a low price given idle spot instances from the cloud.

We allocate tasks according to the following two scheduling policies: (1) Availability-based scheduling: Tasks in the task list will not be assigned to any unavailable GPU instance that is already executing a task. 
(2) Time-based scheduling: 
We estimate the runtime required for each index construction task, and avoid assigning tasks to instances whose remaining active time is insufficient to complete them.
We first sample multiple tiny subsets from the dataset and measure their index construction time. Since construction time scales linearly with dataset size under the same algorithm, hardware configuration and dataset characteristics, we use these results and shard sizes to estimate the build time of larger partitioned data shards under the same indexing settings.
Given this, if we know or the cloud provider notifies that a spot instance will be terminated in several minutes, 
the scheduler prioritizes assigning tasks with estimated run-times less than that to this instance. If no such instances, the scheduler will not assign any task to it.

Lastly, if a GPU instance is unfortunately terminated with a unfinished task running on it, the task scheduler will reallocate this task to another suitable instance in the cloud instance list.

\noindent \textbf{Spot instance cost analysis:} To estimate the cost saving of using spot instances, we design a spot instance cost model. 
Generally, the graph indexing cost of a given dataset can be evaluated by the machine active time multiplied by the machine price. We assume all the spot GPU instances we use have the same price for simple illustration.

In \scaleGANN, throughout the entire index-construction process, the CPU machine remains active, to handle not only shard partitioning and merging, but also shard indexing task scheduling across multiple GPU machines. In contrast, each GPU machine is activated only when it is assigned a shard-index construction task. It is worth noting that multi-GPU parallelism can speed up the overall index-construction time and reduces the CPU-active period. 
However, for GPU cost consumption, multiple cards within the same GPU machine do not incur additional charges, but multiple GPU machine instances running in parallel are billed separately. 
Therefore, the cost of using multiple GPU machines should be calculated as the unit price of a single GPU machine multiplied by the total active time across all GPU machines.
In addition, because we rely on cloud resources, the total machine usage time must also account for the network transfer time for shard data communication between CPU and remote GPUs. 
Overall, the total cost for \scaleGANN if using GPU spot instance(s) is computed as:
(overall construction time + data transfer time) $\times$ CPU price + (aggregated GPU active time + data transfer time) $\times$ GPU spot instance price.

Note that in this work, we apply time-based scheduling policy to guide task assignment which tries to avoid unexpected termination. 
Therefore, the above cost model currently does not consider the cost of spot instance interruption and task rescheduling, and Sec~\ref{sec:experiments} presents a simple cost analysis under this assumption.

\section{Adaptive Vector Partitioning}
\label{sec:partition}

To construct a union and connected index for large dataset, we proposed an adaptive disk-resident data partitioning strategy with selective vector replication for both index quality after merge and build efficiency considering storage and time.

\begin{figure}[t]
    \centering
    \includegraphics[width=0.25\textwidth]{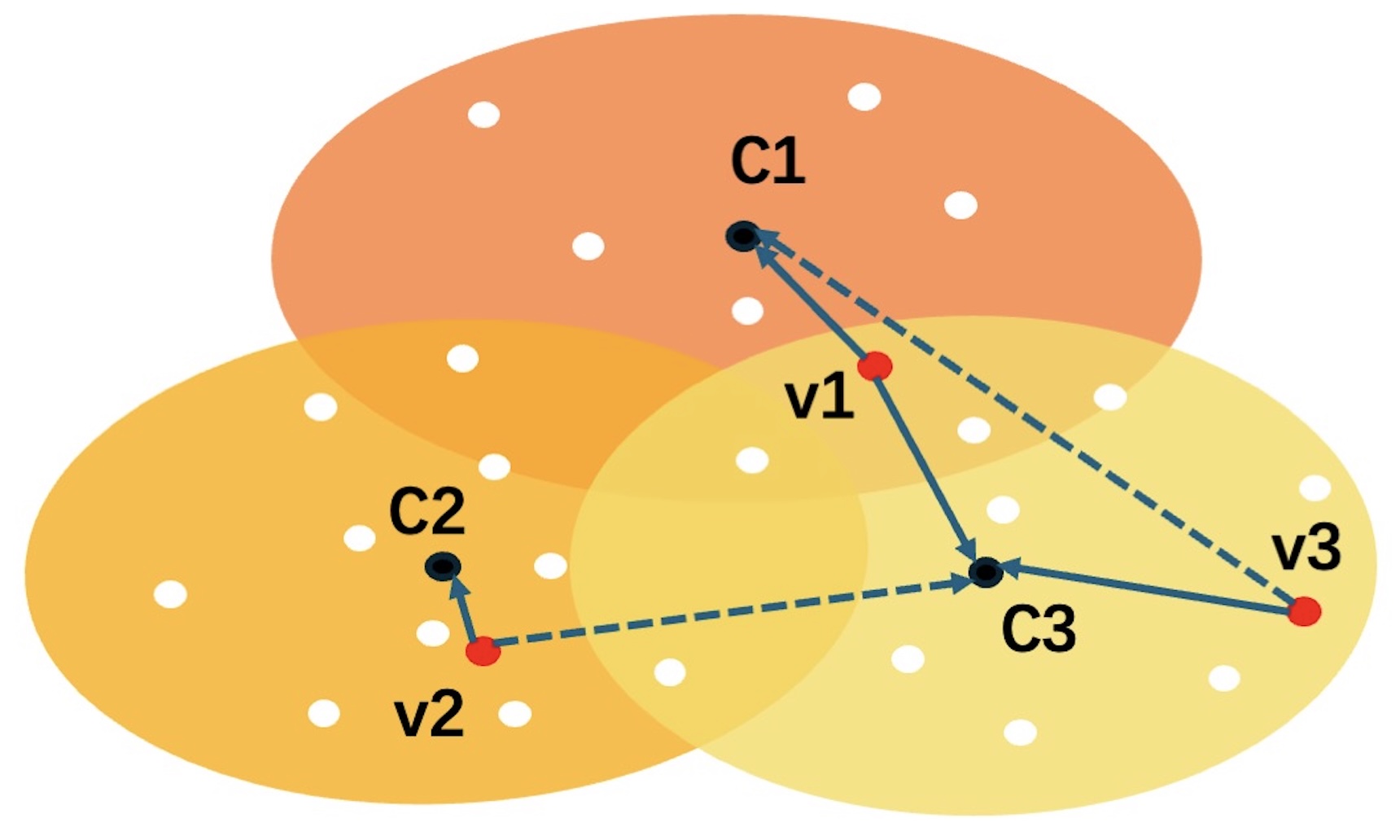}
    \caption{Illustration of Assignment Policy.}
    \vspace{-1em}
    \label{fig:assignment}
\end{figure}

Given limited GPU and CPU memory, prior partition-and-merge-based indexing work partitions the large dataset into multiple smaller data shards using k-means clustering, replicates all the vectors once or more times across partitions to ensure global connectivity, and then builds index on each partition and merge shard indices. 

However, there are two issues. (1) Vectors are loaded and processed block-by-block from disk, and those in earlier blocks may saturate clusters too early, preventing later vectors from being assigned to their nearest clusters. 
To illustrate, we show three cluster centroids $c_1$, $c_2$, $c_3$, and three vectors $v_1$, $v_2$, $v_3$ depicted in Figure~\ref{fig:assignment}, with their respective cluster preferences as follows: ($c_1 > c_3 > c_2$), ($c_2>c_3>c_1$), ($c_3>c_1>c_2$).
Suppose each vector is replicated at least once, with cluster capacity limit 2, and block order $v_1, v_2, v_3$. $v_1$ replicas will be assigned to $c_1$ and $c_3$, then $v_2$ replicas to $c_2$ and $c_3$, saturating cluster $c_3$. As a result, $v_3$ can no longer be fairly assigned to its closest cluster $c_3$, despite being the most appropriate. (2) Since each vector has at least one replica, the dataset size at least doubles, leading to significant memory and processing time overheads.

\subsection{Blockwise-adaptive Assignment}

To ensure fairness in partitioning, we need to mitigate the impact of vector processing order on assignment. We distinguish two types of assignments: assigning (1) an original vector to its nearest available cluster, and (2) its replicas to other clusters. The first guarantees that every vector belongs to at least one cluster, ensuring dataset completeness and locality. The second improves inter-cluster connectivity, but also introduces distant vectors into other clusters. Each cluster must reserve capacity for original vectors processed later for fairness and locality, while also accepting a certain number of replicas for connectivity.

To balance fairness, locality, and connectivity, we introduce a tunable threshold that controls the proportion of cluster space available for replicas. Once this limit is reached, the cluster can only accept original vectors and decline future replicas. Each cluster can have its own replica threshold, adjusted by the data distribution: dense clusters often serve as the nearest choice for many vectors, so they use smaller thresholds to preserve space for unprocessed original vectors.

We further support blockwise runtime adaptive adjustment of thresholds. After each block is processed, we update the data distribution information and thresholds for each cluster based on observed vector assignments. 
Note that during the whole assignment process, to avoid costly disk I/O, the dataset is read only once: for each block, we first assign original vectors to their nearest available clusters, then update cluster distribution statistics and thresholds, finally allocate replicas based on the latest thresholds, and then move to the next block.

\subsection{Selective Replication}\label{ss:replication}
To further reduce data redundancy while maintaining inter-shard connectivity and enhancing shard locality, we introduce a selective replication strategy.

In addition to increased memory and time overhead for storing, manipulating, and indexing the duplicated vectors, full-scale replication introduces redundant assignments. Vectors close to the centroid of their nearest clusters may be unnecessarily replicated to distant clusters, as its neighbors are likely to be only within the cluster and assigning replicas to remote clusters would even create spurious edges. 

To address these issues, we perform only necessary assignments. 
Given vector $v$ and its closest centroid $c$ with distance $d$, for any existing cluster $c'$ with radius $r'$ and distance $d'$ to $v$, the assignment of $v$ to $c'$ is necessary if and only if it obeys two constraints. (1) Distance Constraint: $d' < \epsilon \cdot d$, and (2) Radius Constraint: $d' < \epsilon \cdot r'$, where $\epsilon$ is a tunable parameter.

The intuition is as follows. If $d'$ is much larger than $d$ or $r'$, it can be implied that $v$ is either close to the center of $c$, or far from that of $c'$, rendering the assignment to $c'$ unnecessary.
Based on this strategy, we replicate vectors only to sufficiently close neighboring clusters. 

\begin{algorithm}
\caption{Selective Replica Assignment}
\label{alg:selective-replica-assigment}
\begin{algorithmic}[1]
{\footnotesize 
\Statex \textbf{Input}: Vector block $B$, Number of Clusters
        $k$, Cluster replica threshold $\theta$, Cluster radius $R$, 
        Clusters (set of assigned vectors) $S$,
        $\omega$, $\epsilon$, $\tau$
\Statex \textbf{Output}: Updated clusters $S$}
\vspace{0.5em}
\For{$v \in B$}
    \State $c \gets loadCentroid(v)$ \# original vector assignment
    \State $d \gets dist(v,c)$
    \State $assigned \gets 1$
    \For{$c' \in sorted(Centroids, key=dist(v,c'))$}
        \If{$assigned \geq \omega$} \text{break}
        \EndIf
        \If{$c' \neq c$ \text{and} $checkSizeLimit(c', \theta)$}
            \State $d' \gets dist(v,c')$
            \If{$d' < \epsilon \cdot d$ \text{and} $d' < \epsilon * \tau * R[c']$}
                \State $S_{c'} \gets S_{c'} \cup {v}$
                \State $assigned \gets assigned + 1$
            \EndIf
        \EndIf
    \EndFor
\EndFor
\end{algorithmic}
\end{algorithm}

Algorithm~\ref{alg:selective-replica-assigment} illustrates the assignment of replicas with selective replication. Here $\omega$ denotes the maximum number of clusters a vector can appear in, and $\epsilon$ is a tunable parameter controlling pruning strength. 
For each loaded block $B$, we first record the assignment between each vector and its nearest available cluster (Line 2). We also update each cluster's replica threshold $\theta$ and radius $R$ based on the block data, which serves as inputs for the assignment.

As in Line 5-6, a vector $v$'s assignment ends once all $k$ clusters have been iterated, or cluster replica limit $\omega$ has been reached.  
For each vector $v$ in the current block, we iterate through the clusters in an ascending order of distances from $v$ shown in Line 5. In Line 7-11, we check the size and remaining replica space of the target cluster $c'$, and perform replica assignment from $v$ to $c'$ if it passes both size check and the selective pruning.
During block-by-block manipulation, we observed that in the early stage with fewer processed blocks, the radius of each cluster can be smaller than its actual value. Therefore, we introduce a dynamic radius parameter $\tau$ that is initially large and decreases as the number of processed blocks increases, to remedy the actual radius for a more accurate radius-based pruning as shown in Line 9. 

Recall the example in Figure~\ref{fig:assignment}.
$v_1$ are replicated on $c_1$ and $c3$, while $v_2$ is assigned only to $c_2$, as it lies close to its center. Similarly, $v_3$ is assigned only to $c_3$ because its distance from $c_1$ exceeds the pruning threshold. 
This not only ensures expected assignment of $v_3$ to its nearest cluster $c_3$, it also improves memory and time efficiency during index builds.

\subsection{Parallelism}
We further accelerate the partitioning algorithm using multi-threading on the CPU, primarily in the following three components: (1) The distance computation between vectors and centroids can be executed in parallel. (2) The assignment of multiple vectors can be processed simultaneously. (3) The data distribution of multiple clusters can be updated concurrently.

\tl{To fully achive the potential of parallelism, the key challenge we addressed is maintaining deterministic processing order... Previous work such as DiskANN fail to parallelize ... becasue } Previous work such as DiskANN only leveraged multi-threading for the distance computation between vectors and centroids. One reason for this limitation is that, although using multiple threads can significantly speed up vector assignment and data partitioning, it also results in a non-deterministic processing order. That is, the order in which vectors are processed and written into data shards becomes random, thus the arrangement of vectors within each shard no longer matches the order in the original dataset. 
During shard-level index merging, DiskANN relies on sequential disk reads to efficiently utilize read buffers. If the vector ordering is inconsistent across data shards, it leads to index information being read and merged incorrectly.
To address this, we extend the original implementation by adding a simple buffer state check, enabling us to safely support random disk reads while still maintaining efficient buffer utilization. This ensures that the vector reading order during index merging matches the true original vector order.
\section{Experiments}
\label{sec:experiments}
Our default experimental setup includes a 40-core 80-thread Intel(R) Xeon(R) Gold 6133 CPU @ 2.50GHz with 251G RAM and 2T SanDisk SDSSDH3, and the Tesla V100-SXM2-16GB GPU(s) with a total of 5120 threads. 

In our experiments, we compare four different implementations:
\begin{itemize}
    \item \textbf{\scaleGANN:}
    Our proposed GPU-based system for large-scale ANNS index construction where the dataset is selectively replicated and partitioned, followed by parallel block-wise GPU index construction, and finally merging of shard indexes. Queries are served on CPUs for high throughput and low latency.
    \item \textbf{GGNN:}
    A state-of-the-art GPU-based ANNS system designed for large-scale datasets. It partitions the dataset without replication, constructs a graph index for each block independently, and performs queries on each block separately. Final results are obtained by merging and ranking the per-block query results.
    \item \textbf{Extended CAGRA:}
    A baseline adaptation based on CAGRA, a cutting-edge GPU ANNS system for small datasets. We extend it using GGNN's block-partitioning and block-wise querying approach to enable it to handle large-scale datasets. For each block, Extended CAGRA performs graph index construction and querying, followed by result aggregation and re-ranking.
    \item \textbf{DiskANN:}
    A leading CPU-based ANNS solution for large-scale datasets. It replicates and partitions the dataset, constructs a graph index for each block on CPU, and merges these indexes into a global index. Queries are executed on the merged index.
\end{itemize}

\begin{table}[t]
    \caption{Dataset.}
    \label{tab:dataset}
\vspace{-0.5em}
    \centering
    \footnotesize
    \begin{tabular}{l|l|l|l}
    \hline
      & \textbf{Size} & \textbf{Dimension} & \textbf{Data Type} \\
     \hline
     \textbf{Sift100M} & 100,000,000 & 128 & uint8 \\
     \hline
     \textbf{Deep100M} & 100,000,000 & 96 & float \\
     \hline
     \textbf{MicrosoftTuring100M} & 100,000,000 & 100 & float \\
     \hline
     \textbf{Laion100M} & 100,000,000 & 768 & float \\
     \hline
     \textbf{Sift1B} & 1000,000,000 & 128 & uint8 \\
     \hline
\end{tabular}
\end{table}

The datasets used in our experiments are listed in Table~\ref{tab:dataset}. Sift, Deep, and SimSearchNet (with query sets) are from the BIGANN benchmarks~\cite{bigann-benchmark}. Laion100M is a sampled subset of Laion5B~\cite{laion5B,laion5B-downlad}, and the query set is created by sampling 10,000 items from Laion5B, following VectorDBBench~\cite{VectorDBBench}.
In principle, our framework is capable of scaling to the full Laion5B dataset, provided sufficient disk space is available. However, due to the computational demands and hardware limitations, processing the entire dataset would require prohibitively long runtimes on our current setup. Therefore, for experimental analysis, we sample 100M vectors, which already requires 2.5 times more raw storage than Sift1B.
More importantly, since index construction time scales linearly with dataset size ~\cite{CVPR_tutorial, ootomo2024cagra, groh2022ggnn},
results on 100M-scale datasets can serve as reliable estimates for billion-scale performance.

Codes and results are available in an anonymous repo ~\cite{ScaleGANN_Code_Repo}.

\subsection{Overall Result}
In this section, we illustrate the effectiveness of our partition with selective replication based on \scaleGANN and the other partition-and-merge approach DiskANN. 
Then we present the build and search results of four benchmarks, and comprehensively analyze \scaleGANN's performance in both indexing and querying.

\subsubsection{Selective Vector Partitioning}
\label{sssec:selectivity}

\begin{table}[t]
\caption{Sift100M Index Time (s) at Different Selectivity $\epsilon$.}
\label{tab:epsilon-study}
    \centering
    \footnotesize
    \begin{tabular}{l|l|l|l|l}
    \hline
      \textbf{\scaleGANN} & \textbf{$\epsilon=1.1$} & \textbf{$\epsilon=1.2$} & \textbf{$\epsilon=1.5$} & \textbf{Original} \\
     \hline
     \textbf{Proportion} & 33.3\% & 54.3\% & 81.8\% & 100\% \\
     \hline
     \textbf{Overall Time (s)} & 4216 & 4726 & 3428 & 5837 \\
     \hline 
     \textbf{Build-Only Time (s)} & 2220 & 2570 & 3025 & 3564 \\
     \hline 
\end{tabular}
\end{table}

Compared to DiskANN's default setting where each vector is replicated once, we show a wise choice of selectivity $\epsilon$ can reduce 20\%-70\% vector replication, thus saving disk and memory usage, reducing data processing and indexing time.
While accelerating the index build, the search quality is maintained or even improved instead of trading off one for the other.

\begin{figure}[t]
    \centering
    \includegraphics[width=0.45\textwidth]{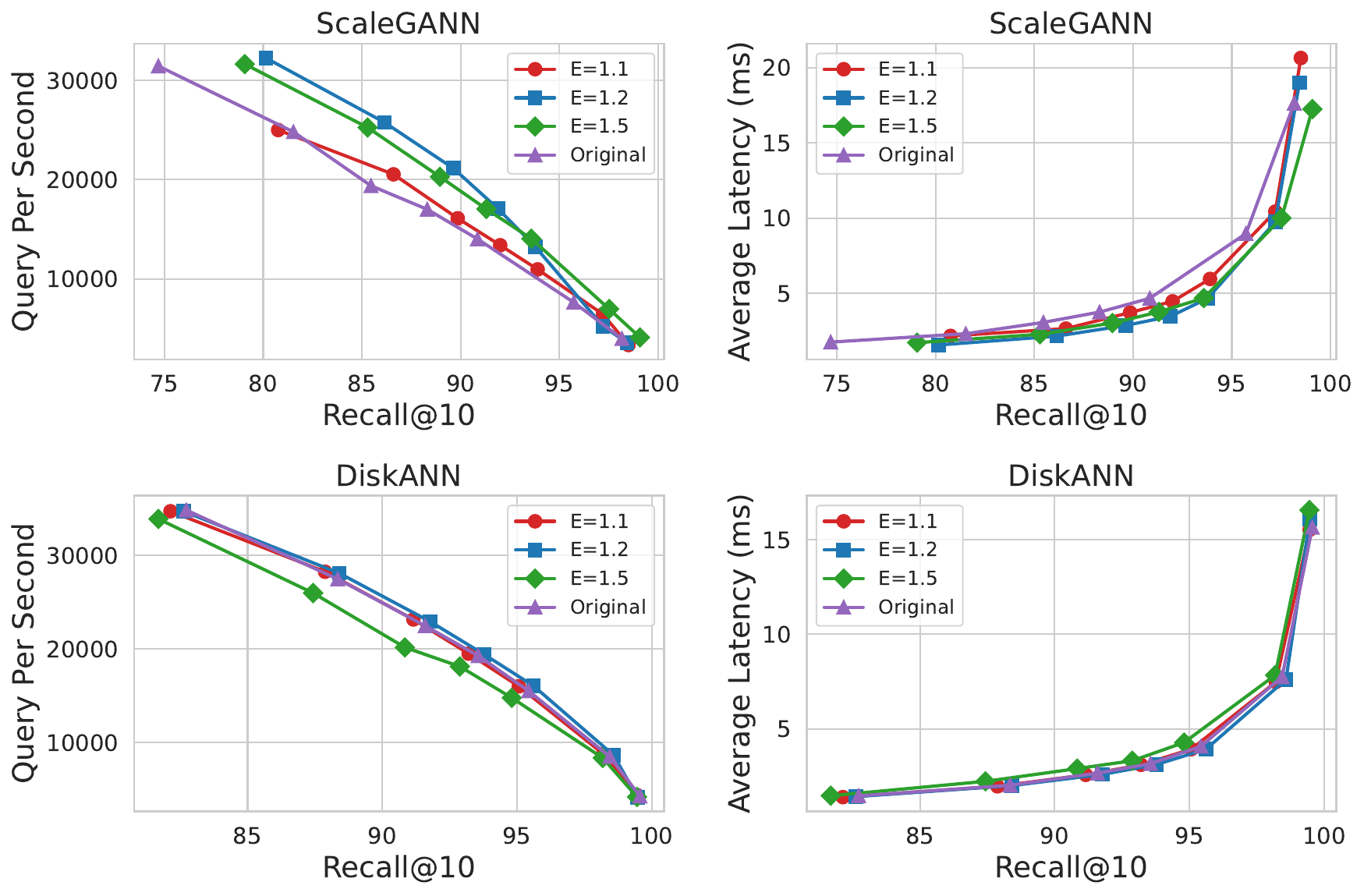}
    \vspace{-1em}
    \caption{Sift100M Search Quality at Different Selectivity $\epsilon$.}
    \label{fig:search-epsilon}
    \vspace{-0.5em}
\end{figure}

Table ~\ref{tab:epsilon-study} presents \scaleGANN's build performance under different selectivity $\epsilon$. "Proportion" denotes the percentage of replicated input vectors, while "Overall Time" and "Build-Only Time" represent the total indexing time and shard indexing time respectively (excluding both partitioning and merging).
In addition, Figure~\ref{fig:search-epsilon} illustrates \scaleGANN's corresponding query performance under varying duplication rates. 

As the replication proportion decreases, we can save more index build time. In particular, "Build-Only Time" shows an even near-linear reduction corresponding to the decrease in overall data size.
Interestingly, a moderate choice of $\epsilon$ can maintain or even improve query quality at the same time. 
At $\epsilon=1.1$ with a replica reduction of up to 66.7\%, both search latency and query per second are improved, due to the elimination of assigning distant vectors to clusters during replication explained in Section~\ref{ss:replication}. 

Besides, when applying this approach to DiskANN’s Vamana index, the conclusion still holds, which is also shown in Figure ~\ref{fig:search-epsilon}. This further demonstrates the generality of our selective duplication across different indexing algorithms.

\subsubsection{Indexing Time and Quality}
\label{sssec:build}

\begin{table}[t]
    \caption{Index Construction Time (s).}
    \label{tab:index-build-time-multi-datasets}
\vspace{-0.5em}
    \centering
    \footnotesize
    \resizebox{1.01\linewidth}{!}{
    \begin{tabular}{l|l|l|l|l|l}
    \hline
      \multicolumn{2}{l|}{\textbf{Dataset}} & \textbf{Extended CAGRA} & \textbf{GGNN} & \textbf{DiskANN} & \textbf{\scaleGANN} \\
     \hline
     \multicolumn{1}{l}{\textbf{Sift100M}} & Overall & 1790 & 1287 & 9597 & 4727 \\
     \multicolumn{1}{l}{} & Build-Only & 1693 & 1287 & 7848 & 2570 \\
     \hline 
     \multicolumn{1}{l}{\textbf{Deep100M}} & Overall & 1775 & 2004 & 21159 & 5222 \\
     \multicolumn{1}{l}{} & Build-Only & 1673 & 1840 & 19155 & 2797 \\
     \hline
     \multicolumn{1}{l}{\textbf{MSTuring}} & Overall & 2063 & 10977 & 20221 & 6672 \\
     \multicolumn{1}{l}{\textbf{100M}} & Build-Only & 1871 & 10744 & 18089 & 3480 \\
     \hline 
     \multicolumn{1}{l}{\textbf{Laion100M}} & Overall & 5483 & 44735 & 62109 & 11259 \\
     \multicolumn{1}{l}{} & Build-Only & 3850 & 43131 & 57163 & 6504\\
     \hline 
     \multicolumn{1}{l}{\textbf{Sift1B}} & Overall & 18155 & 13899 & 119039 & 70617 \\
     \multicolumn{1}{l}{} & Build-Only & 17362 & 13025 & 83732 & 27676 \\
     \hline
\end{tabular}}
\end{table}

\begin{figure}[t]
    \centering
    \includegraphics[width=0.48\textwidth]{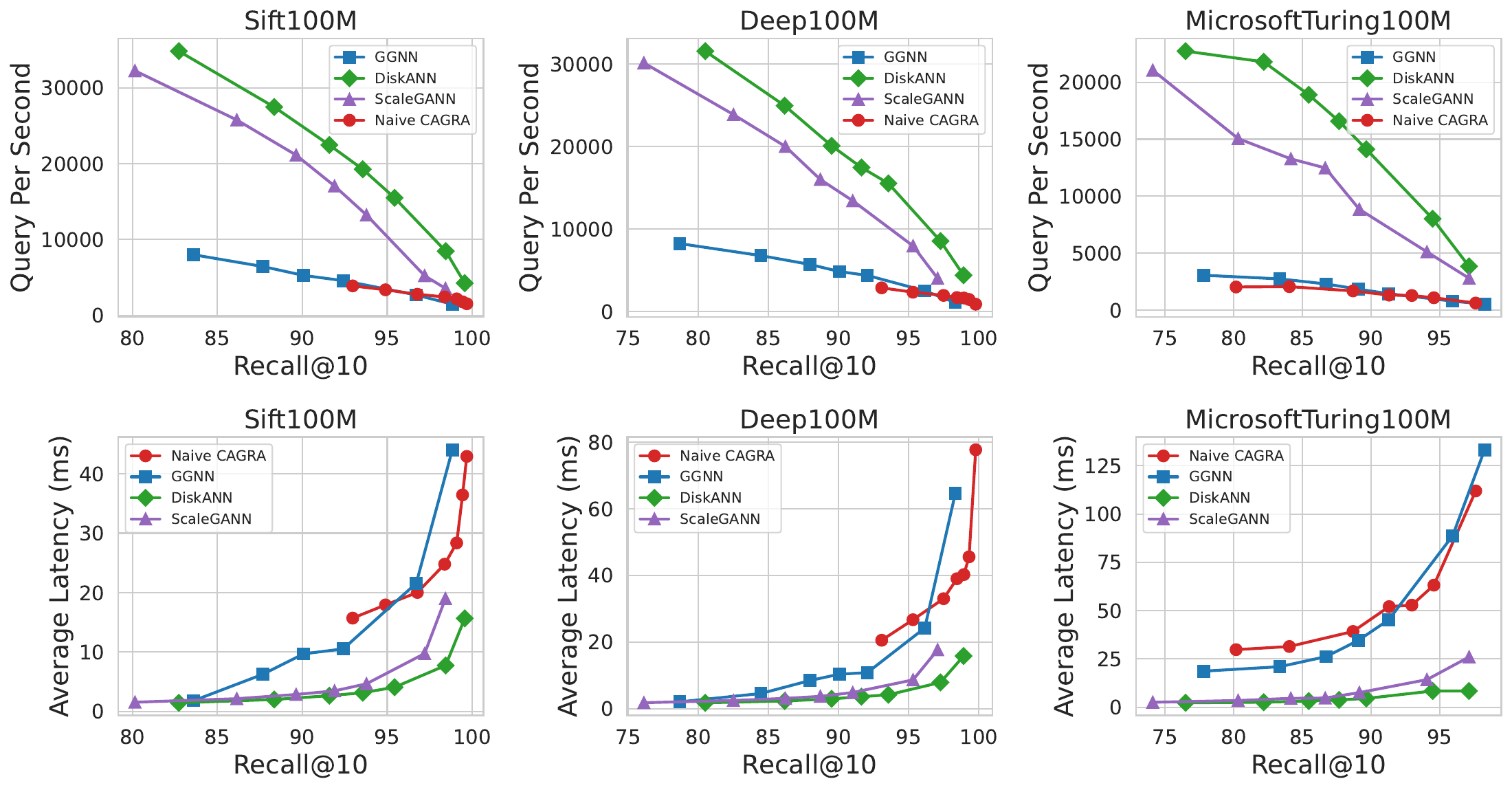}
    \caption{Search Performance of Low-dimensional Datasets.}
    \label{fig:search-small-dataset}
\end{figure}

\begin{figure}[t]
    \centering
    \includegraphics[width=0.48\textwidth]{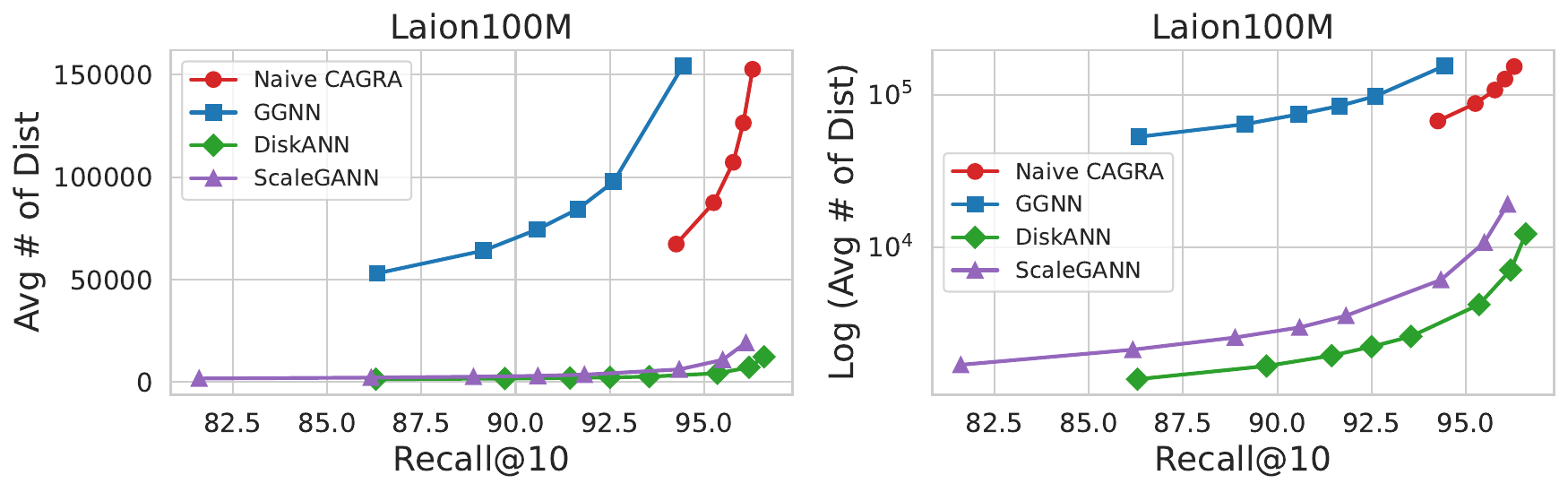}
    \caption{Search Performance of Laion100M.}
    \label{fig:search-laion}
\end{figure}

We compare the index construction time of all the benchmarks on a single machine (CPU/GPU) with a 16 GB memory budget, as shown in Table~\ref{tab:index-build-time-multi-datasets}.
Corresponding search efficiency of these indexes are provided in Figure~\ref{fig:search-small-dataset} and Figure~\ref{fig:search-laion}, fairly based on a unified CPU query algorithm following DiskANN’s search strategy.

In Table~\ref{tab:index-build-time-multi-datasets}, the default build degree and intermediate build degree are 64 and 128, which is the widely adopted setting for large datasets. For GGNN, we adopt a build degree of 20 for Sift and 24 for Deep, following the settings in its paper.
During partition, DiskANN and \scaleGANN allows one replication for a vector while \scaleGANN uses a selectivity factor of $\epsilon=1.2$ to remove unnecessary replicas. For Extended CAGRA and GGNN, datasets are naively splitted without vector replication. 
We propose two metrics "Overall Time" and "Build-Only Time" for a comprehensive performance comparison.
For DiskANN and \scaleGANN, the overall index construction time includes data partitioning, shard index construction and index merging. While for CAGRA and GGNN, only partitioning and shard indexing are considered in overall time.
Besides, the "Build-Only" times represent the shard indexing time only for all the approaches, excluding both partitioning and merging.

For search experiments, due to limited space, we compare only the top-10 recall with query per second (QPS) and average latency as in Figure~\ref{fig:search-small-dataset}.
For high-dimensional large-scale Laion100M, we perform a disk-based search in Figure~\ref{fig:search-laion}, and use the average number of distance computed as a proportional proxy for both QPS and latency. 
Though we could compress Laion100M using vector quantizations and thus enable in-memory search, search optimization is not the focus of this work. 
Therefore, we leave it for future search enhancements.

\textbf{\scaleGANN vs Extended CAGRA and GGNN.} We first compare the partition-and-merge based \scaleGANN with split-only Extended CAGRA and GGNN, where all approaches use GPU for index construction acceleration. With selective vector replication and index merging, \scaleGANN substantially improves search performance without introducing prohibitive index construction overhead. 

Both Extended CAGRA and GGNN exhibit higher search latency and significantly lower QPS than \scaleGANN at equivalent recall. For example, as in Figure~\ref{fig:search-small-dataset}, at 95\% recall, the two split-only methods incur over 3× higher latency and achieve roughly only 1/3 the QPS of the split-and-merge methods. The results on Laion100M in Figure~\ref{fig:search-laion} confirm the same trend. 
Note that for split-only methods, though GPU-based querying can potentially boost the QPS, latency and excessive vector reads still remain to be bottlenecks without index merging.

\scaleGANN's indexing time is better than GGNN if at the same build degrees, with up to 4x overall improvement on Laion100M. However, GGNN can be faster on low-dimensional datasets like Sift100M, Deep100M and Sift1B, where GGNN builds the graph index using small build degrees (20 and 24, respectively). Notably, if both were configured with the same build degree of 64, GGNN's build time would increase by nearly 5×, making it slower than \scaleGANN.

Compared to Extended CAGRA, \scaleGANN requires roughly 2–3× the total construction time, mainly because the dataset is duplicated to ensure shard connectivity, effectively doubling its size and processing time, along with additional disk-based complex data partitioning and index merging. This overhead is magnified in Sift1B given increased slow disk operations. However, since \scaleGANN performs a selective replication to to minimize vector replication, the build-only time which reflects only the graph index construction time on GPU is always less than 2× that of CAGRA. This strategy also reduces the time of partitioning and merging to a certain extent, benefited by saved computation, disk read and write.

\textbf{\scaleGANN vs DiskANN.} We then compare the GPU-based \scaleGANN with CPU-based DiskANN, both adopting the split-and-merge strategy.

Compared to DiskANN, \scaleGANN achieves an significant overall speedup in index construction, and up to 5.5× overall acceleration on high-dimensional Laion100M. This improvement stems from two key factors: (1) the use of GPU acceleration, which significantly improves distance computation efficiency in high-dimensional spaces, and (2) our selective replication strategy, which reduces unnecessary data duplication to lower data processing and indexing overhead. 
Meanwhile, since both DiskANN and \scaleGANN rely on CPU and costly disk-based implementations for partitioning and merging large datasets, which incur similar and non-negligible overhead towards the "overall" time. Therefore, if isolating the shard index construction stage, the actual index build-only time speedup of \scaleGANN over DiskANN increases to almost 9x. 

Importantly, \scaleGANN is anticipated to maintain its advantage on other billion-scale datasets instead of just Sift1B, since the "build-only" indexing time scales approximately linearly with dataset size. Though the "overall" time of \scaleGANN indicates potential improvement space for disk-based data processing procedures in the future.

Besides, though under the current settings, \scaleGANN's search performance is slightly worse than DiskANN’s, this is mainly due to differences between the indexing algorithms, instead of trading-off the search performance for construction time. DiskANN uses Vamana graph, while CAGRA builds KNN graphs. However, this is orthogonal to our framework which actually can integrate with any of these indexing methods.
In fact, search performance of the same given index can even be improved by adjusting \scaleGANN’s parameters, especially the selectivity factor $\epsilon$ which should be tuned for different datasets and is further explored in Sec ~\ref{sssec:selectivity}.





\subsection{Results under Various Settings}

\subsubsection{Increased Build Degree}

\begin{table}[t]
    \caption{Overall Construction Time (s) of Laion100M.}
    \label{tab:build-laion100M}
    \vspace{-0.5em}
    \centering
    \footnotesize
    \resizebox{1.01\linewidth}{!}{
    \begin{tabular}{l|l|l|l|l}
    \hline
      \textbf{Graph Degree} & \textbf{Extended CAGRA} & \textbf{GGNN} & \textbf{DiskANN} & \textbf{\scaleGANN} \\
     \hline
     \textbf{R=32, L=64} & 5022 & 20309 & 29309 & 9477 \\
     \hline
     \textbf{R=64, L=128} & 5483 & 44735 & 62109 & 11259 \\
     \hline
     \textbf{R=128, L=256} & 7142 & 49708 & 149065 & 16311 \\
     \hline
\end{tabular}
}
\end{table}

According to Table~\ref{tab:build-laion100M}, we observe an increase on the overall index construction time as build degree becomes large, though the growth rate varies across different methods considering their different data processing overhead. 
Under a larger index degree, the GPU based approaches can bring larger acceleration due to the increased proportion of distance computation tasks.
At the same time, \scaleGANN with selective replication maintains its affordable replication overhead ($\sim$2x) compared to Extended CAGRA, and performs better than GGNN.
On Laion100M, for instance, when the build degree is set to 128 and the intermediate degree is set to 256, \scaleGANN achieves up to 9× overall speedup over DiskANN, and 3x over GGNN.


\subsubsection{Multi-GPU parallelism}

\begin{table}[t]
    \caption{Index Build-Only Time (s) over GPU Parallelism.}
    \label{tab:gpu-parallel}
    \vspace{-0.5em}
    \centering
    \footnotesize
    \begin{tabular}{l|l|l|l|l}
    \hline
      & \textbf{Sift100M} & \textbf{Deep100M} & \textbf{MSTuring100M} & \textbf{Laion100M} \\
     \hline
     \textbf{1 GPU} & 2570 & 2797 & 3480 & 6504 \\
     \hline
     \textbf{2 GPU} & 1495 & 1557 & 1992 & 3394 \\
     \hline
     \textbf{4 GPU} & 877 & 882 & 1158 & 2003 \\
     \hline 
\end{tabular}
\end{table}

Multi-GPU parallelism can further accelerate index construction. As shown in Table ~\ref{tab:gpu-parallel}, using 2 and 4 V100 GPUs achieves near-linear speedup for shard index build stage compared to a single GPU under the same build parameters $R=64$, $L=128$ and $\epsilon=1.2$.
Note that speedup is not exactly linear due to system overhead and uneven shard sizes.
However, if with larger datasets and more shards, task distribution could become more balanced, and the acceleration could become ideal scaling.

This confirms that our graph index construction framework which divides the workload into many small and independent tasks is highly suitable for multi-GPU parallelism.
Each GPU will independently process one or more data shards, while each separate shard construction takes only a few minutes.
For example, with Sift100M, the dataset is divided into 16 shards, each taking approximately 160 seconds to build. In this case, if using 4 GPUs, each GPU handles 4 shards in average, leading to efficient parallel construction. 
This design further makes our system naturally adaptable to cloud-native spot instance environments where small and fast tasks can be flexibly scheduled across dynamic cheap GPU resources.

\subsection{Spot Instance Cost Analysis}

Lastly, we provide a simple cost analysis of DiskANN and \scaleGANN based on the cost model in Sec~\ref{sec:framework}. In this study, DiskANN uses a regular CPU machine with similar conditions to our local machine, while \scaleGANN uses one GPU spot instance (with 4 V100 GPU cards) for shard index construction and the same CPU machine for partitioning and merging. 


According to AWS ECS~\cite{awsspot}, a regular Linux CPU machine with around 80 threads, 200G RAM and 2T disk (e.g., c5d.24xlarge) is at \$3.9-4.6/h, while a Linux GPU machine with 4 16G V100 (e.g., p3.8xlarge) has regular price \$13.7/h and spot instance price changing normally between \$1.22-3.67/h. 
While all the instances offer at least 10Gbps network bandwidth, we estimate the total data transfer time as: number of shards $\times$ 16GB / network bandwidth. This is because each shard transfer task consists of shard data sent to GPU and the index returned to CPU, which involves at most 16GB data bounded by the GPU memory.

Using Laion100M index construction in Table~\ref{tab:index-build-time-multi-datasets} as an example, it generates less than 100 shards, indicating a 1600G upper bound for data transfer amount and correspondingly at most 160s, namely 0.045h, data transfer time. 
Besides, we observe that DiskANN's overall build time is 17.25h in Table~\ref{tab:index-build-time-multi-datasets}, while Laion100M takes 0.56h to build the index using 4 V100 parallelism in Table~\ref{tab:gpu-parallel}. Then we also obtain \scaleGANN's partition and merge time calculated by its overall time minus the index build-only time, which is 1.32h as shown in Table~\ref{tab:index-build-time-comparison}. Thus, the overall build time of \scaleGANN is 1.88h (0.56h+1.32h). Therefore, the cost estimate for DiskANN with regular CPU is at least \$67.3 ($17.25 \times 3.9$), while \scaleGANN costs at most \$11.1 ($(1.88+0.045) \times 4.6 + (0.56+0.045) \times 3.67$) which is even 6x cheaper than DiskANN.

Notably, in this scenario, even on-demand cloud GPUs can be more cost-effective than CPUs. This is primarily due to the efficiency of GPU acceleration on high-dimensional datasets, which substantially reduces runtime and, consequently, the overall cost. For low- and mid-dimensional datasets, using GPU spot instances still offers indexing speedups while greatly lowering the cost.
Moreover, although older regular GPUs are already relatively affordable, newer GPUs with larger memory and more threads enable faster and more stable indexing by reducing the number of shards and speeding up distance computations. In such cases, spot instances of the latest GPUs can achieve better performance at the same or even lower cost than older regular GPUs.

\section{Related Work}

\stitle{Storage Management}
To serve large-scale vector indexing and search, DiskANN~\cite{jayaram2019diskann} offloads data to disk and preserves 10-20\% memory for efficiency. Starling~\cite{wang2024starling} further reorders the disk layout to improve locality and reduce search disk access.
Besides, LM-DiskANN~\cite{pan2023lm} and AiSAQ~\cite{tatsuno2024aisaq} propose disk-only designs that do not rely on memory in extreme senarios. 
Alternative ANNS systems including SPANN~\cite{chen2021spann} and FusionANNS~\cite{tian2025fusionanns} adopt IVF indexes for disk-resilient design. 
In particular, FusionANNS optimize index build on CPUs by reducing random I/O via layout and I/O de-duplication.
Furthermore, some ANNS systems offload data to emerging memory devices such as PMEM~\cite{ren2020hm}, CXL~\cite{jang2023cxl}, UPMEM PIM~\cite{chen2024memanns}, and SmartSSD~\cite{tian2024scalable}. 
All are orthogonal to our study.

\stitle{Acceleration}
Distance calculation can be accelerated through CPU parallelism (e.g., DiskANN ~\cite{jayaram2019diskann}) or GPU parallelism (e.g., GANNS~\cite{yu2022gpu} and CAGRA~\cite{ootomo2024cagra}), though the latter is constrained by memory of a single GPU. Besides, asynchronous GPU data transfer help hide transffering time~\cite{bhatotia2012shredder, yang2024seraph},
while quantization techniques ~\cite{wang2025accelerating} (e.g. PQ, SQ, and PCA) further reduce memory footprint and trade some accuracy for speed.
For large datasets across multiple machines, methods explore partitioning, multi-shard parallelism, and load balancing: examples include GPU-based GGNN, CPU-based Starling~\cite{wang2024starling}, and distributed shard builds in SOGAIC~\cite{shi2025scalable}. \scaleGANN adopts a more flexible, cost-efficient approach by leveraging GPU spot instances in parallel while maintaining a CPU union index to balance efficiency and cost.

\stitle{Graph Update} Streaming scenarios with frequent vector updates (insertions and deletions) motivate specialized ANNS systems: SPFresh~\cite{xu2023spfresh} extends SPANN, while IP-DiskANN~\cite{xu2025place} and FreshDiskANN~\cite{singh2021freshdiskann} enhance DiskANN with update capabilities.
However, as changes and updates accumulate over time, index rebuilding is still necessary to maintain search quality, especially for graph-based indices.



\stitle{Systems with spot instances}
Spot instances are widely employed on various workloads for cost-effectiveness, such as ML training~\cite{duan2024parcae}, MapReduce tasks~\cite{chohan2010see}, and MPI applications~\cite{taifi2011spotmpi}.
Specifically, recent AI serving systems like SkyServe~\cite{mao2025skyserve} and SpotServe~\cite{miao2024spotserve} consider spot GPU instances, with LLM specific parallelization configurations. 
Meanwhile, mechanisms~\cite{danysz2020aws,wang2018empirical,kim2024making} of check-pointing, migration and node replicas are proposed to handle unexpected terminations of spot instances.
To the best of our knowledge, \scaleGANN is the first ANNS graph indexing system to exploit spot GPU instances for both cost efficiency and performance.

\section{Conclusion}

In this paper, we propose an end-to-end GPU-based cloud-native ANNS graph index construction framework, \scaleGANN, that is highly efficient, scalable, and cost-effective. 
We enhance \scaleGANN's disk-resident extendable partition-and-merge strategy with adaptive vector assignment and selective replication, optimizing time and storage efficiency while preserving the overall search quality. 
In addition, \scaleGANN's framework with a novel resource allocation strategy achieves both efficiency and cost-effectiveness. 
We use spot GPUs exclusively for the one-time computation-intensive shard indexing tasks which are also compatible to multi-GPU parallelism, while still leaving partitioning, merging and long-term querying on CPUs.
Lastly, while our framework can be integrated with any graph index build algorithm, in this paper we apply CAGRA to our implementation. Experiment results show that our approach can bring up to 9x acceleration at a even lower cost compared with DiskANN, while maintaining similar search QPS and latency at the same recall.

Our work also lays the foundation for several valuable future directions.
A natural extension is to address the instability of spot instances by developing more robust task recovery mechanisms. 
For example, live migrating unfinished tasks to another instance, or designing checkpoint-based task resuming methods to avoid full re-execution upon interruption.
Additionally, our current setup assumes homogeneous spot instance types,
Therefore, it opens up opportunities for specific designs for task re-allocation and cost analysis under heterogeneous settings with available instances varying in GPU memory size and compute capacity. 
Lastly, our approach demonstrates how cost-effective, pay-as-you-go cloud resources can be leveraged to complete one-time, compute-intensive ANNS index construction, encouraging further exploration of cloud-native or serverless solutions for scalable vector search.

\bibliographystyle{IEEEtran}
\bibliography{short}

\end{document}